\documentclass[%
 aip,
rsi,%
amsmath,amssymb,%
 reprint,%
]{revtex4-1}
\usepackage{graphicx}% Include figure files
\usepackage{dcolumn}% Align table columns on decimal point
\usepackage{bm}% bold math
\newcommand{\R}{\mathbb{R}}

\newcommand{\N}{\mathbb{N}}
\newcommand{\Z}{\mathbb{Z}}
\newcommand{\T}{\mathbb{T}}
\newcommand{\kd}{K}
\newcommand{\rmd}{\mathrm{d}}
\newcommand{\rme}{\mathrm{e}}

\newcommand{\gr}{g_\mathrm{r}}
\begin{document}

\preprint{AIP/123-QED}

\title{Pinning and Depinning: from periodic to chaotic and random media}% Force line breaks with \\
\thanks{Research supported through grant NSF DMS--1612441.}

\author{N. Ankney}
 \affiliation{Department of Mathematics, Michigan State University, 619 Red Cedar Road, East Lansing, MI 48808}%Lines break automatically or can be forced with \\
\author{M. Avery}
 %\homepage{http://www.umn.edu/~scheel.}
\affiliation{%
School of Mathematics, University of Minnesota, 206 Church St, Minneapolis, MN 55455, USA%\\This line break forced% with \\
}
\author{T. Khain}%
 %\email{Second.Author@institution.edu.}
\affiliation{ 
Department of Mathematics, University of Michigan, 530 Church St, Ann Arbor, MI 48109%\\This line break forced with \textbackslash\textbackslash
}%
\author{A. Scheel}
\email{scheel@umn.edu}
 \homepage{http://www.umn.edu/~scheel.}
\affiliation{%
School of Mathematics, University of Minnesota, 206 Church St, Minneapolis, MN 55455, USA%\\This line break forced% with \\
}%

\date{\today}% It is always \today, today,

\begin{abstract}
We study propagation of dissipative structures in inhomogeneous media with a focus on pinning and depinning transitions. We model spatial complexity in the medium as generated by dynamical systems. We are thus able to capture transitions from periodic to quasiperiodic, to homoclinic and heteroclinic, and to chaotic media. Depinning bifurcations exhibit universal laws depending on extreme value statistics that are encoded in the dimension of ergodic measures, only. A key condition limiting this approach bounds spatial Lyapunov exponents in terms of interface localization and we explore the breakdown of smoothness and universality when this condition is violated and fluctuations in the medium occur on length scales shorter than a typical interface width.
\end{abstract}

%Use showkeys class option if keyword
\maketitle

\begin{quotation} 
We analyze front propagation in a simple bistable lattice differential equation in an inhomogeneous medium. When the space dependence of the medium is modeled over a smooth manifold and is stationary with respect to lattice translations, we argue that a simple reduced dynamical system captures all key features of the dynamics. In particular, we predict universal depinning exponents in terms of statistical properties of a generating diffeomorphism acting on the model manifold. We present computational tools that exploit this reduced description to predict locations and scaling laws of depinning transitions. Finally, we show both in chaotic media as well as in a simple toy heteroclinic medium how results break down when Lyapunov exponents of the generating diffeomorphism are comparable to the exponential localization of the front interface. 
\end{quotation}

%%%%%%%%%%%%%%%%%%%%%%%%%%%%%%%%%%%%%%%%%%%%%%%%%%%%%%%%%%%%%
\section{\label{sec:intro}Introduction}
%For cat map we want the <x,y> values at the edge of the map corresponding to smallest xi so we can mod back to 0 (assuming we want to do that)
%We need to resolve the definition of xi with the infinite domain issue

Dynamics in spatially extended bistable systems are often mediated by the propagation of an interface. For gradient flows, the propagation of the interface dissipates energy as the energetically favorable of the two stable states spreads in the medium. The relation between the speed and the difference in energies of the two stable states gives a kinetic relation that captures the essential feature of such a spatially extended system. Front dynamics in inhomogeneous media can be significantly more complex. If one defines an average energy difference $\mu$ between the two stable states, and measures average speeds in terms of this average difference, speeds turn out to be not strictly monotone, even in very simple systems. The prime reason for these more complicated kinetic relations is the possibility of pinning, where a ``local'' energy difference may be negative and therefore prevent front propagation, although the average energy difference is positive. In periodic media and in lattices, this phenomenon is fairly well understood \cite{erneux,clerc,pomeau}, with applications including signal propagation in myocardial tissue \cite{keener,keener2}, martensitic phase transitions  \cite{james_pinning,bhattacharya_pinning}, or chemical reaction fronts \cite{Laplante1992}. In these situations, depinning, that is, small non-zero speed past a non-zero critical value of the average energy difference, occurs when two pinned states disappear in a saddle-node bifurcation. Speeds $s$ near this depinning transition scale as $s\sim(\delta\mu)^{1/2}$, due to the standard scaling of passage times near a saddle-node \cite{erneux}. Our interest here is in pinning and depinning in more complex media, striving to recover some universality in asymptotics of speeds near depinning, and studying the breakdown of these universal laws. 

\begin{figure*}[t!]
\includegraphics[width=180mm,trim={3.5cm 0 3.5cm 0},clip]{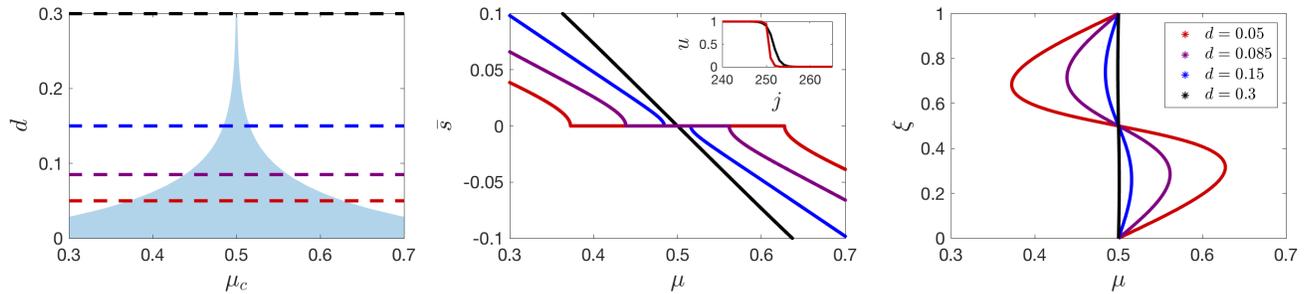}
% \caption{The depinning transition ($\mu_c$) as a function of $d$, the coupling constant, in a constant medium. The internal periodicity of the lattice creates a non-zero pinning region; as $d$ is increased, the pinning region shrinks. In the continuous case ($d \to \infty$), pinning occurs at only $\mu_c = 0.5$.}
\caption{Pinning regions in the $(\mu,d)$--plane (shaded, left) with kinetic relations $\bar{s}(\mu;d)$ (center) and pinning relations $\xi(\mu;d)$ (right) for selected $d$ values. The inset (center) shows the widening of the front interface with increasing $d=0.05\ldots 0.3$.   }\label{fig:lattice_d}
\end{figure*}

To fix ideas, we consider the  Nagumo (or Allen-Cahn) equation in an inhomogeneous lattice
\begin{equation}\label{e:lat}
u_{j}'=d(u_{j+1}+u_{j-1} - 2 u_j) + u_j(1-u_j)(u_j-a_j-\mu),
\end{equation}
with initial conditions $u_j\to 1$, $j\to -\infty$, and $u_j\to 0$, $j\to\infty$. For $a_j\equiv 0$, pinning regions $|\mu|<\mu_\mathrm{c}(d)$ decrease exponentially in $d$ for large $d$; see Figure \ref{fig:lattice_d} for kinetic relations, pinning relations, and pinned regions. 

Our goal is to demonstrate that in the framework of ``dynamically generated media'', one can systematically study kinetic relations and pinning relations, and, in particular, uncover universality in depinning asymptotics. The class of dynamically generated media that we focus on is modeled through actions of diffeomorphisms $\psi$ on smooth manifolds $\mathcal{M}$ as 
\[
a_j=A\circ\psi^j(m),\quad \psi:\mathcal{M}\to\mathcal{M},\quad A:\mathcal{M}\to (0,1).
\]
In this sense, each point $m\in \mathcal{M}$ is associated with a medium $\{a_j\}_{j\in\Z}$, and we refer to $m$ as a realization of the medium.

The first examples of interest are  quasiperiodic media $\mathcal{M}=\T^\kd=\R^\kd/\Z^\kd$ with rationally independent frequencies $(\omega_j)_{1\leq j\leq \kd}$, e.g.
\[
\psi(m)=m+\omega\, \mathrm{mod}\, 1,\quad A(m)= \sum_{j=1}^\kd \alpha_j \cos(2\pi\omega_j m_j).
\]
From here, we can then envision increasing the complexity in the medium through changes in the diffeomorphism $\psi$. The second example we consider here is a chaotic medium modeled over $\T^2$ with $\psi$ the cat map
\[
\psi(x,y)=(2x+y,x+y) \, \mathrm{mod}\,  1.
\]
We shall explain below how different choices of functions $A$ and choices of initial conditions that are ``typical'' for different ergodic measures can lead to a variety of statistical behaviors. Depinning asymptotics are, however, governed by a simple statistical property of the medium, the dimension of the ergodic measure $\kappa$,
\begin{equation}\label{e:dpasy}
\bar{s}(\mu)\sim (\mu-\mu_\mathrm{c})^{\frac{1-\kappa}{2}},\ \kappa<1,\qquad \bar{s}(\mu)\sim 1,\ \kappa>1,
\end{equation}
where $\bar{s}$ is a suitably defined average speed of propagation.
We compare these behaviors with some simple choices of random media, for which one can readily intuit depinning asymptotics. 

These simple asymptotics break down when fluctuations in the medium occur on small spatial scales, compared to length scales associated with the front interface. We quantify this limit and demonstrate why and how our simple description breaks down near these boundaries. In the following, we first describe more precisely how we model inhomogeneous media and front dynamics in such media. We illustrate this point of view, first  in the example of a quasi-periodic medium and then for a chaotic medium. We compare with different types of random media and finally explore limits of this point of view.

%%%%%%%%%%%%%%%%%%%%%%%%%%%%%%%%%%%%%%%%%%%%%%%%%%%%%%%%%%%%%%%%%%%%%%%%%%%%%%%%%

\section{Dynamically generated media and front dynamics on a medium}
\begin{figure}[h!]
\includegraphics[width=75mm]{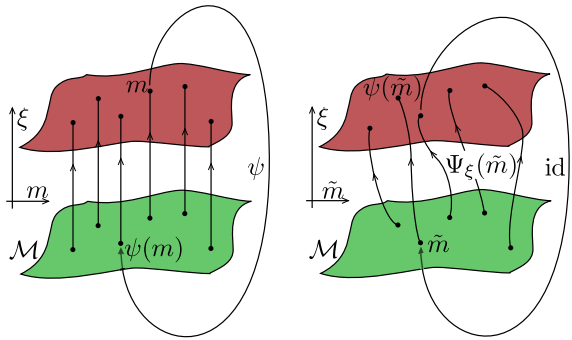}
\caption{Schematic plot of dynamics on $\tilde{\mathcal{M}}$ in steady frame \eqref{e:pm} (left) and comoving frame \eqref{e:comov} (right).}
\label{fig:compact}
\end{figure}

The simplest description of front dynamics tracks the position of the front for a given realization of the medium $m$, that is, for coefficients generated through $a_j=A\circ\psi^j(m)$. In the case of the lattice Nagumo equation, the position $\xi$ can be defined for instance through 
\[
\xi(u)=\sum_{j=-J}^\infty u_j,
\]
as long as the front interface is near a location $j_0$ with $j_0-J\gg 1$. For our purposes equivalent definitions could rely on solving 
\[
\sum_{j\in\Z} S(j-\xi)u_j=1/2, 
\]
for a strongly localized $S$ with $\int S=1$. We suppose that front dynamics in a given medium can effectively be reduced to dynamics of the position $\xi$,  
\begin{equation}\label{e:pm}
\xi'=s(\xi;m),\quad m'=0,
\end{equation}
and $s(\xi;m)$ is a speed, depending, of course, on the specific definition of position $\xi(u)$. Choices of position function $\xi$ such that $\xi(\{u_j\})=\xi(\{u_{j-1}\})+1$ guarantee that dynamics at front position $\xi+1$ are the same as dynamics at position $\xi$, changing the realization of the medium to $\psi(m)$, that is, 
\[
s(\xi;\psi(m))=s(\xi+1;m)
\]
Note also that average speeds, defined as times to propagate one lattice site, $s_1:=(T_1-T_0)^{-1}$, where $\xi(T_1)=\xi(T_0)+1$, do not depend on the definition of $\xi$. We can therefore study front dynamics relative to the lattice (and the medium), considering \eqref{e:pm} on 
\begin{equation}\label{e:q}
(\xi,m)\in\tilde{\mathcal{M}}=[0,1]\times \mathcal{M}/\sim
\end{equation}
with identification $(1,m)\sim (0,\psi(m))$; see Figure \ref{fig:compact}.
The gluing of $\xi=1$ to $\xi=0$ can sometimes be trivialized, when  $\psi$ is homotopic to the identity through $\Psi_t$, 
\[
\psi=\Psi_1,\ \mathrm{id}=\Psi_0,
\ \partial_t\Psi_t(m)=h(\Psi_t(m),t), 0\leq t\leq 1.
\]
We introduce new coordinates through $(\xi,m)=(\xi,\Psi_{\xi}(\tilde{m})$, trivializing the identification $(1,\tilde{m})\sim (0,\tilde{m})$ with the new differential equation 
\begin{equation}\label{e:comov}
\xi'=\tilde{s}(\xi;\tilde{m}),\qquad \tilde{m}'=\tilde{s}(\xi;\tilde{m})\cdot\tilde{h}(\xi,\tilde{m}),
\end{equation}
where $\tilde{s}=s(\xi;\Psi_\xi(\tilde{m}))$, $\tilde{h}=h(\Psi_\xi(\tilde{m}),\xi)$. 
The new medium $\tilde{m}$ can be thought of as the medium in a comoving frame; compare Figure \ref{fig:compact} for the two geometric view points.

For quasiperiodic media, one chooses $\Psi_\xi(m)=m+\omega\xi$, $h=\tilde{h}=\omega$, and finds
\begin{equation}\label{e:qp}
\xi'=s(\xi,\tilde{m}+\omega\xi),\qquad 
\tilde{m}'=-s(\xi,\tilde{m}+\omega\xi)\cdot\omega.
\end{equation}
The resulting vector field can be easily computed and we illustrate the result below  for the systems with a periodic $\kd=0$ and with a quasi-periodic medium $\kd=1$. 

Since the flow in both \eqref{e:pm} or \eqref{e:comov} is determined primarily by the scalar field $s$, one is first interested in zeros of $s$ which we shall refer to as pinned states. When we include the dependence of $s$ on the parameter $\mu$, we can typically solve $s(\xi;m,\mu)=0$ for $\mu$ and obtain the pinning relations $\mu_\mathrm{p}(\xi,m)$, that give the applied force $\mu$ that forces a pinned state at position $\xi$ in the medium $m$. Extrema of $\mu$ determine depinning transitions, at least when $\psi$ is transitive on $\mathcal{M}$, that is, when trajectories of $\psi$ explore all possible values of $m$. We show pinning relations and their correspondence with average speeds and pinning regions in Figure \ref{fig:lattice_d}, and give more examples below. 

Pinning relations also give good estimates on barriers to propagation when noise is added to the system. Potential differences between stable and unstable pinned states, visible here as areas between the pinning relation and a $\mu\equiv const$ level set, are known as Peierls-Nabarro barriers and determine rates for jumps of the interface between lattice sites.

%%%%%%%%%%%%%%%%%%%%%%%%%%%%%%%%%%%%%%%%%%%%%%%%%%%%%%%%%%%%%%%%%%%%%%%%%%%%%
\section{Ergodicity and depinning asymptotics}\label{s:3}

In the framework of \eqref{e:pm} on the quotient manifold \eqref{e:q}, we can now study front dynamics from a dynamical systems point of view. Pinned front locations are equilibria. For $P$-periodic media, $\mathcal{M}=\{m_0\}$, $P\in\N$, we obtain dynamics on the circle $\xi\in [0,P]/\sim$, $\xi'=s(\xi)$. Dynamics are either periodic when $s\neq 0$, or converge to pinned states. Average speeds are easily obtained as a harmonic average by integrating $\frac{1}{P}\int_0^P s(\zeta)^{-1}\rmd\zeta=\bar{s}^{-1}$. Near depinning, speeds are close to zero near a degenerate equilibrium, $\mu=\mu_\mathrm{c}+\tilde{\mu}$, $\xi=\xi_\mathrm{c}+\tilde{\xi}$, 
\begin{equation}\label{e:snp}
\tilde{\xi}'\sim c_0\mu+c_2\tilde{\xi}^2\end{equation}
and the harmonic average is at leading order given by the integral 
\begin{equation}\label{e:pdp}
\bar{s}\sim \left(\int_{|\tilde{\xi}|\leq\delta}\frac{1}{c_0\tilde{\mu}+c_2 \xi^2} \right)^{-1} \sim \pi^{-1}\sqrt{c_0c_2\tilde{\mu}}.
\end{equation}
In general, the dynamics of the medium can of course be arbitrarily complex. 
We then solve \eqref{e:pm} by computing the time for $\xi$ to increase from $0$ to $1$, and then average times over the iterates $m_j=\psi^j(m)$ of the medium
\[
T_1(m)=\int_0^1\frac{\rmd\zeta}{s(\zeta;m)},\ \bar{s}=\left(\lim_{N\to\infty}\frac{1}{N}\sum_{j=1}^{N}T_1(m_j)\right)^{-1}.
\]
If the initial condition is typical for an ergodic measure $\rmd\nu(m)$ on $\mathcal{M}$  associated with $\psi$ acting  on $\mathcal{M}$, this sum of $T_1$ along orbits can be replaced by a phase space average using Birkhoff's ergodic theorem
\[
\bar{s}=\left(\int_\mathcal{M} \rmd\nu T_1(m) \right)^{-1}=\left(\int_\mathcal{M} \rmd\nu \int_0^1\frac{\rmd\zeta}{s(\zeta;m)}  \right)^{-1}
\]
In quasi-periodic media, all realizations are typical for Lebesgue measure, and the integral is simply the $\kd+1$-dimensional Lebesgue measure. 

When $s(\xi;m)$ almost vanishes for some near-critical parameter value $\mu=\mu_\mathrm{c}+\tilde{\mu}$,  and near a non-degenerate minimum $\xi=\xi_\mathrm{c}+\tilde{\xi}$, $m=m_\mathrm{c}+\tilde{m}$, the integral is at leading order
\begin{equation}\label{e:sasy}
\bar{s}\sim\left(
\int_{|\tilde{m}|+|\tilde{\xi}|\leq \delta} 
     \frac{\rmd\tilde{\zeta}\rmd\nu }{c_0\tilde{\mu}+c_2[\zeta,m]}  
\right)^{-1}
\end{equation}
for some bilinear positive definite form $c_2$. A simple scaling analysis shows that 
\begin{equation}\label{e:asy}
\bar{s}(\mu_\mathrm{c}+\tilde{\mu})\sim\left\{\begin{array}{ll}
1, & \kappa>1,\\
-\log|\tilde{\mu}|^{-1},& \kappa=1,\\
\tilde{\mu}^{\frac{1-\kappa}{2}},& \kappa<1,
\end{array}\right.
\end{equation}
where $\kappa$ is the dimension of the ergodic measure $\nu$, that is, the measure of a ball of size $r$ scales as $\nu(B_r)\sim r^\kappa$. Similar asymptotics have been derived for the case of continuous media $x\in\R$. \cite{AScheel1}

For quasi-periodic media with more than one frequency, $\kd>1$, we always see the first case, hard depinning, with discontinuous speeds near the transition. For quasiperiodic media with one frequency (irrational relative to the lattice periodicity), one can be slightly more precise and compute the coefficient at leading order
\begin{equation}\label{e:qpexp}
\bar{s}(\mu_\mathrm{c}+\tilde{\mu})=-\frac{D}{2\pi}\log|\tilde{\mu}|^{-1},
\end{equation}
where $D=\sqrt{c_2^{11}c_2^{22}-\frac{1}{4}c_2^{12}c_2^{12}}$ is the geometric  (Gaussian) average of the  curvatures of $s$ at the zero, where $c_2[\tilde{\xi},\tilde{m}]=c_2^{11} \tilde{\xi}^2+c_2^{12}\tilde{\xi}\tilde{m}+c_2^{22} \tilde{m}^2$. 

We will demonstrate below that coefficients in these asymptotics can be readily obtained numerically with very high accuracy using direct Newton solvers.
%%%%%%%%%%%%%%%%%%%%%%%%%%%%%%%%%%%%%%%%%%%%%%%%%%%%%%%%%%%%%%%%%%%%%%%%%%%%%%%%%%%%%
\section{\label{sec:coherent}Coherent Media}

\begin{figure}[h!]
\includegraphics[width=90mm]{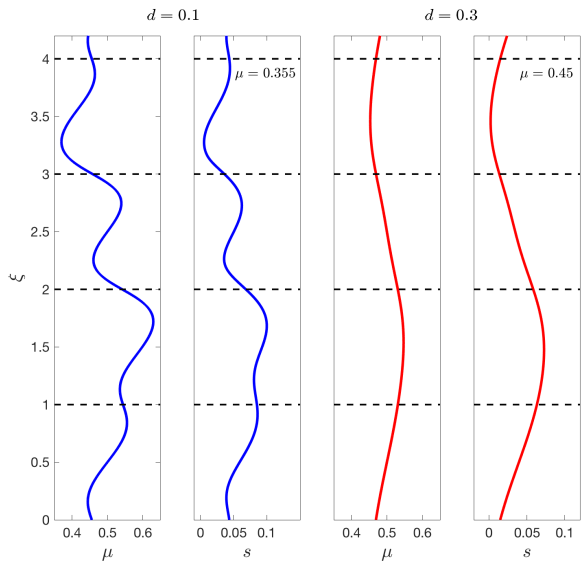}
\caption{Pinning relations $\mu_\mathrm{p}(\xi)$ and instantaneous speeds $s(\xi)$ for a periodic medium, $a_j = 0.1\cos(\frac{\pi}{2}j)$. For small $d$ (left panel), the lattice periodicity emerges to add a superimposed period-1 pattern to the medium. For larger $d$, approximating a continuous medium (right panel), this small-scale periodicity disappears. Chopping the plots along the horizontal dashed lines yields pinning relations $\mu_\mathrm{p}(\xi;m)$ and instantaneous speeds $s(\xi;m)$ as functions of $\xi\in[0,1]$ and the medium $m\in \{0,1,2,3\}$  with identification $(\xi,m)=(1,j)=(0,j+1)$.}
\label{fig:periodic}
\end{figure}

We apply and illustrate the analysis in the case of periodic media ($\kd=0$) and quasiperiodic media ($\kd=1$). We first consider $a_j=0.1 \cos(\pi j/2)$ of period 4, with $d=0.1$ and $d=0.3$. We can consider the front dynamics on $\xi\in \R/(4\Z)$, with trivial medium $\mathcal{M}_1=\{0\}$, or front dynamics on $\xi\in \R/(\Z)$, with medium $\mathcal{M}_4=\{0,1,2,3\}$, $\psi(j)=j+1\,\mathrm{mod}\,4$. Figure \ref{fig:periodic} shows pinning relations and effective dynamics on $\xi\in \R/(4\Z)$, with horizontal cuts showing how these dynamics would decompose into dynamics on $\mathcal{M}_4$. 
 Pinning relations yield critical values $\mu_\mathrm{c}$ and depinning asymptotics according to \eqref{e:pdp} can be readily verified; see Figure \ref{fig:lattice_d} for an illustration of the square root asymptotics for average speeds past the last pinned state.

\begin{figure}[h!]
\includegraphics[width=\columnwidth]{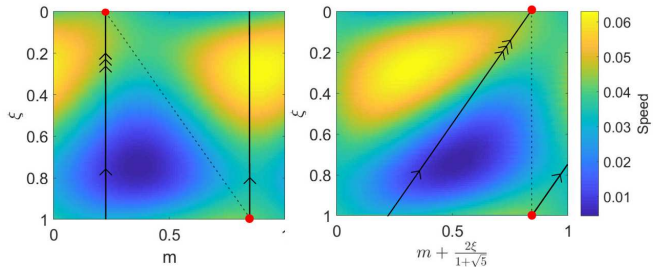}
\caption{Interface speeds $s(\xi,m)$ from \eqref{e:pm} (left) and in comoving medium $\tilde{s}(\xi;m)$ \eqref{e:qp} computed for $\mu = 0.36$, with medium $a = 0.1\cos(2\pi(\gr j + m))$.}
\label{fig:quasi_phi_speed}
\end{figure}

In quasiperiodic media, we consider $a_j = 0.1\cos(2\pi(\gr j + m))$, $\gr=(1 + \sqrt{5})/2$. Computing the derivative of the position $\xi=\sum_{j\geq J} u_j$, we find the speed $s(\xi;m)$, from \eqref{e:pm} which can then be transformed to the speed in a comoving medium \eqref{e:qp}; see Figure \ref{fig:quasi_phi_speed}. 
 
\begin{figure}[h!]
\includegraphics[width=\columnwidth]{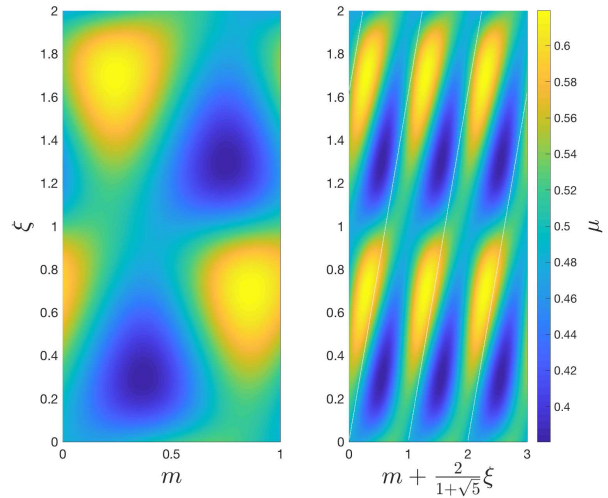}
\caption{Contour plots of pinning relations $\mu(\xi;m)$, for a quasiperiodic medium of the form $a_j = 0.1\cos\left(2\pi(\gr j + m)\right)$. The shear transformation to a comoving medium $m \mapsto m + \gr^{-1}\xi$ creates a pinning surface periodic both in $\xi$ and $m$. Maxima and minima of $\mu$ mark critical depinning values $\mu_\mathrm{c} = 0.3804, 0.6196$.}
\label{fig:quasi_pinned}
\end{figure}

Equilibria in $\eqref{e:pm}$ can be calculated directly. We continued the branch of equilibria in $\mu$ and $m$, tracking the interface location $\xi$, and pinning relations, $\mu=\mu_\mathrm{p}(\xi;m)$ as a function of  $m$ and $\xi$; see Figure \ref{fig:quasi_pinned}.

\begin{figure}[h!]
\includegraphics[width=\columnwidth,trim={1cm 0 2.5cm 0},clip]{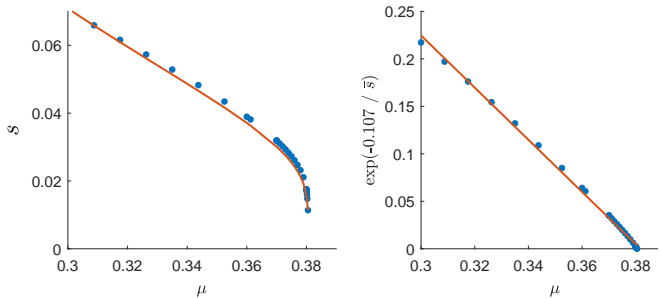}
\caption{Average speeds $\bar{s}(\mu)$ for a  medium $a = 0.05\cos(2\pi(\gr j + m))$, compared to predictions from \eqref{e:qppred}. The critical value of $\mu$ agrees to 4 digits.}
\label{fig:quasi_per_fits}
\end{figure}

We computed extrema of $\mu_\mathrm{p}(\xi;m)$ using a Newton method. We first define equations for equilibria and their position, with independent variables $(u_j)_{|j|\leq J},\xi,m,\mu$,
\begin{align}
F(\{u_j\},m)_j=
d(u_{j+1}-2u_j+u_{j-1})+f(u_j;m)&=0\notag\\
\sum_j u_j-\xi&=0\label{e:nqp}
\end{align}
with Neumann boundary conditions and where 
\[
f(u_j;m)=u_j(1-u_j)\left(u_j-\mu-0.1\cos(2\pi(\gr j + m))\right).
\]
\begin{figure}[h!]
\includegraphics[width=95mm]{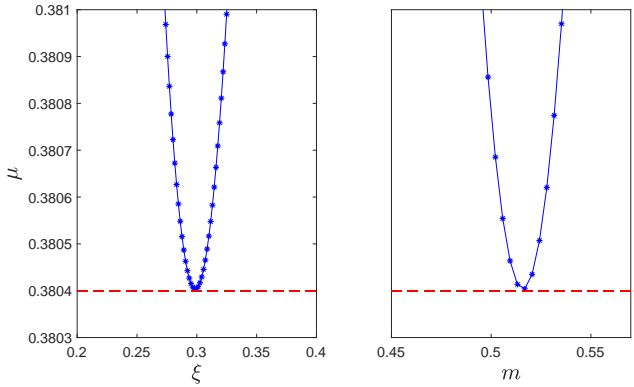}
\caption{Pinning relations near the minimum of $\mu_\mathrm{p}(\xi,m)$, showing both precise values of the depinning transition and curvature near the minimum. }
\label{fig:quasi_depin}
\end{figure}
We then solve these equations with the additional condition that derivatives of $\mu$ with respect to $m$ and $\xi$ vanish. From the resulting system, we directly obtain the principal curvatures at the extrema. Figure \ref{fig:quasi_depin} illustrates the pinning relations showing cross sections through the surface of pinned states near the minimum.
We can then use  \eqref{e:qpexp} to predict speed asymptotics near the depinning transition as
\begin{equation}\label{e:qppred}
\bar{s}\sim - 0.107 (\log|\mu-\mu_\mathrm{c}|)^{-1}, \ \mu_\mathrm{c}=0.3804. 
\end{equation} Figure \ref{fig:quasi_per_fits} shows speeds measured in direct simulations using Matlab's \textsc{ode45} integrator for times $T\sim 10^{7}$ with appropriate shifts in a system of size $1000$, compared to our prediction.
The techniques presented here can readily predict critical depinning parameter values $\mu_\mathrm{c}$ for quasiperiodic media with more than one additional frequency. Depinning is, however, hard, with discontinuous $\bar{s}(\mu)$ at $\mu=\mu_\mathrm{c}$ such that asymptotics are not meaningful at leading order. 

%%%%%%%%
\section{\label{sec:chaotic}Chaotic Media}
Chaotic dynamics of $\psi$ offer a rich set of examples of media as we shall demonstrate now, bridging the spectrum from simple periodic media over quasiperiodic media to chaotic, unstructured media.  We shall focus here on a rather explicitly understood example of a chaotic map that will nevertheless serve as an example of a variety of phenomena. Consider therefore the cat map, an Anosov diffeomorphism of $\T^2$ given through
\[
\left(\begin{array}{c}m_x\\m_y\end{array}\right)\stackrel{\psi}{\longrightarrow} \left(\begin{array}{cc}2&1\\ 1&1\end{array}\right) \left(\begin{array}{cc}m_x\\m_y\end{array}\right) \,\mathrm{mod}\,1.
\]
The map $\psi$ is area preserving and in fact ergodic with respect to Lebesgue measure \cite{ce}. On the other hand, all rational points $(m_x,m_y)$ are periodic, with associated ergodic Dirac measures, and there are a plethora of explicit homoclinic and heteroclinic trajectories between periodic orbits, giving rise to shift dynamics and associated invariant Bernoulli measures.\cite{ce} Regardless of the ergodic measure that captures the statistics of the trajectory associated with a particular medium, we can study dynamics of \eqref{e:pm} on $[0,1]\times \T^2/\sim$, with identification \eqref{e:q} given now by the cat map. Since the cat map is not isotopic to the identity (it acts nontrivially on the fundamental group), it is not possible to construct a comoving medium in the sense of \eqref{e:comov}.

\begin{figure}[t!]
\includegraphics[width=\columnwidth]{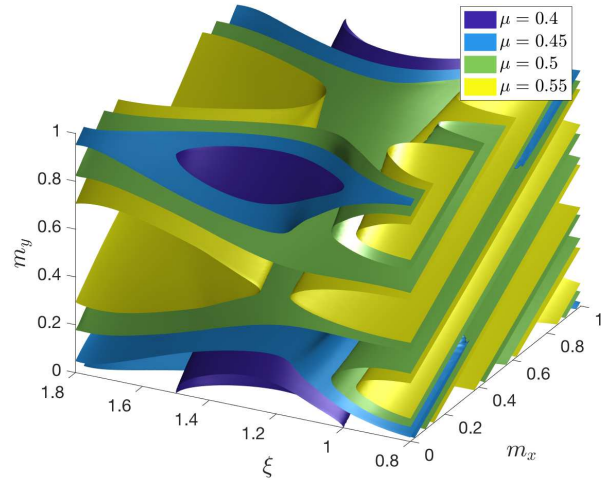}
\caption{The isosurfaces of the pinned states in the cat map medium for four values of $\mu$ with $d=0.1$, $\epsilon=0.1$. Surfaces show locations $\xi$ and realizations of the medium $(m_x,m_y)$ at which pinned states exist at a particular value of $\mu$. }
\label{fig:cat_iso}
\end{figure}

\begin{figure}[t!]
\includegraphics[width=0.99\columnwidth]{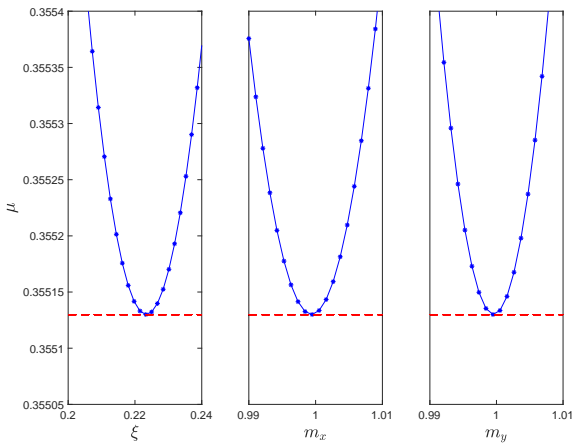}
\caption{Cross sections of the pinning relations $\mu_\mathrm{p}$ showing the non-degenerate minimum inside the dark blue bubble of Figure \ref{fig:cat_iso} at $\xi=0.22$, $m_x=m_y=0$. }
\label{fig:cat_depin}
\end{figure}

\begin{figure}[t!]
\includegraphics[width=0.99\columnwidth,trim={.5cm 0 .5cm 0},clip]{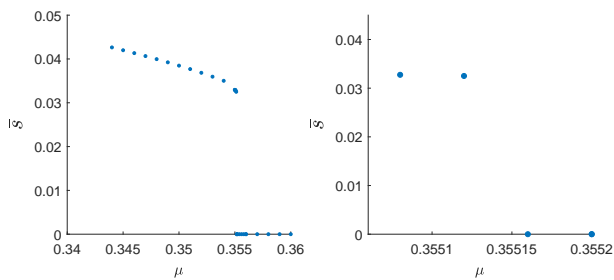}
\caption{Measured average speeds in the cat map medium for Lebesgue-typical initial conditions, with detail near the depinning transition.}
\label{fig:depin_cat}
\end{figure}

We computed pinning relations $\mu_\mathrm{p}(\xi,m_x,m_y)$, for the choice $a_j=\epsilon \cos(2\pi m_{x,j})$; see Figure \ref{fig:cat_iso}. Note that the surfaces are periodic in $m_x$ and $m_y$ but periodic in $\xi$ with the identification $(1,\,m_x,\,m_y)\sim (0,\,2m_x+m_y,\,m_x+m_y)$. We found a minimum of $\mu_\mathrm{p}(\xi,m_x,m_y)$ at $\mu_\mathrm{c}=0.3551$; see Figure \ref{fig:cat_depin}. Experimental speed measurements for a medium generated with initial condition $m_x =1/\sqrt{2}, m_y = 1/\sqrt{5}$ over time intervals $T\sim 10^7$ confirm this depinning transition as a discontinuous transition with critical $\mu$ agreeing at $10^{-4}$ accuracy; see Figure \ref{fig:depin_cat}.

Non-Lebesgue typical initial conditions lead to dramatically different depinning transitions. Initial conditions on the period-3 orbit $p_0=(\frac{1}{2},\frac{1}{2})\to p_1=(\frac{1}{2},0)\to p_2=(0,\frac{1}{2})\to(\frac{1}{2},\frac{1}{2})$, are of course ergodic with respect to Dirac measures at the periodic orbit, and sample the pinning relations along $\xi\in[0,1)$, $m=p_0,p_1,p_2$, that is, along lines parallel to the $\xi$-axis with intercepts $p_j$ at $\xi=0$ in Figure \ref{fig:cat_iso}. One finds depinning transitions at minimal $\mu_\mathrm{c}=0.3852$ and maximal $\mu_\mathrm{c}=0.6311$, with expected square-root asymptotics of the speed. 

\begin{figure}[t!]
\includegraphics[width=75mm]{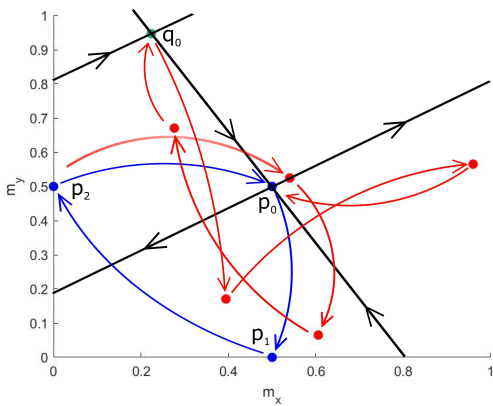}
\caption{Homoclinic orbit to the period-3 orbit in the cat map.}
\label{f:hetcat}
\end{figure}

\begin{figure}[h!]
\includegraphics[width=90mm]{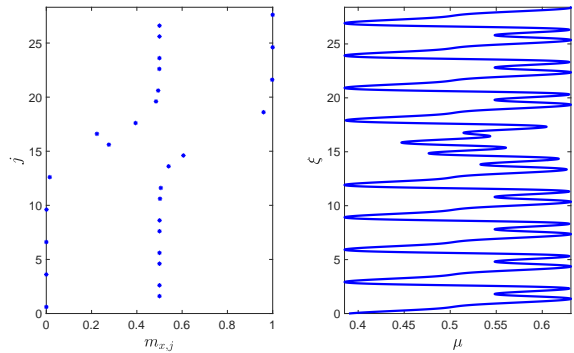}
\caption{The medium $m_{x,j} = \psi^j(m)$ (left) and pinning relations (right) along the homoclinic orbit.}
\label{f:hetpin}
\end{figure}

More interesting are orbits of the cat map that are homoclinic to $\{p_0,p_1,p_2\}$. Since stable and unstable foliations are explicitly given by the eigenvectors of $\psi$ on the covering space $\R^2$, we can explicitly determine a primary homoclinic point $q_0=\frac{1}{2\sqrt{5}}(1 + 2\sqrt{5},2 + \sqrt{5})\mod 1$, such that 
\[
\psi^{3j}(q_0)\to p_0,\quad |j|\to\infty.
\]
Figure \ref{f:hetcat} shows the homoclinic orbit on the torus, and Figure \ref{f:hetpin} shows pinning relations relative to the trajectory of the medium. 

We find that pinning is stronger near the periodic orbit and weaker during the excursion. 

Orbits that stay in a small vicinity of the homoclinic orbit can be described by concatenations of homoclinic excursions and periodic loops. Hyperbolic shadowing guarantees that a formal infinite patching of sequences $\alpha=\{\psi^j(q_0),-3\leq j\leq 2\}$ and $\beta=\{p_0,p_1,p_2\}$ actually corresponds to trajectories that follow the sequence itinerary quite closely.\footnote{Extending the homoclinic sequence beyond length 6 would make this patching more accurate but phenomena of interest here more subtle since pinning is strongest near the periodic points.} Media of the form $(m_j)=(\ldots\alpha\alpha\beta\alpha\beta\beta\ldots)\in \Sigma_2=\{\alpha,\beta\}^\Z$ are therefore included in our cat map medium, for a variety of measures such as those induced by Bernoulli measures on two-symbol sequences. 

\begin{figure}[h!]
\includegraphics[width=\columnwidth,trim={.cm 0 .cm 0},clip]{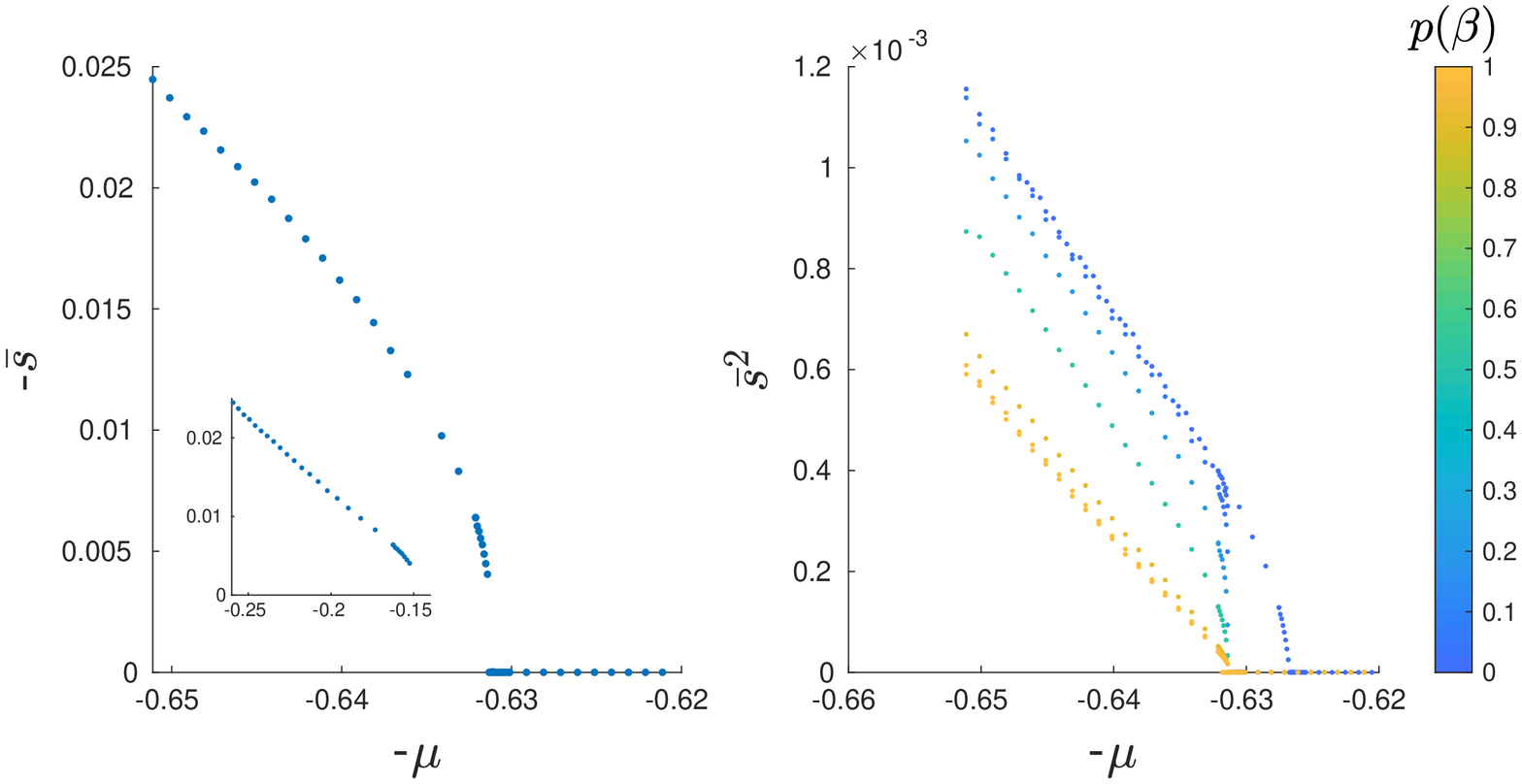}
\caption{Depinning in the cat map for quasiperiodic trajectories near the homoclinic (left) and for trajectories according to a Bernoulli measure near the homoclinic (right);  inset (left) shows best predicted logarithmic fit  $(\frac{-1}{\log{|\mu - \mu_\mathrm{c}|}},-\bar{s})$.}
\label{fig:catbernoulli}
\end{figure}

We investigated depinning numerically for quasiperiodic sequences choosing the block $\alpha$ at the $\ell$'th instance  whenever $\ell\gr \,\mathrm{mod}\,1\leq \gr-1$, and the block $\beta$, otherwise. We found depinning at $\mu_\mathrm{c}=0.6313$, between the critical $\mu$ values of periodic and homoclinic excursion, as expected, with logarithmic depinning asymptotics. 
We also constructed sequences according to Bernoulli measures, choosing $\beta$ with probability $p$ and $\alpha$ otherwise; see Figure \ref{fig:catbernoulli}. We found good fits $\bar{s}\sim \tilde{\mu}^{1/2}$, which would be consistent with a zero-dimensional ergodic measure $\kd=0$. We shall give heuristics for these asymptotics in the next section; compare \eqref{e:markovspeed}.

\section{\label{sec:random}Random Media}
\begin{figure}[h!]
\includegraphics[width=\columnwidth]{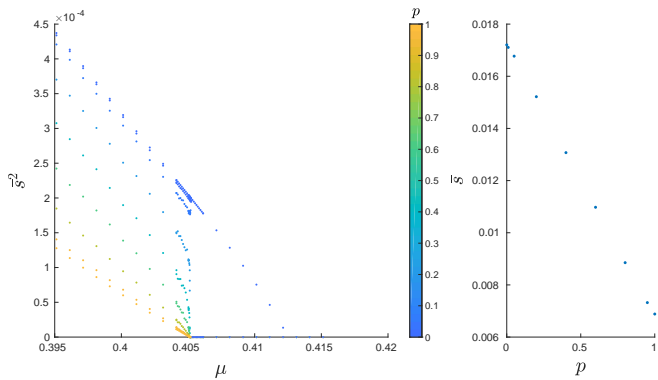}
\caption{Depinning asymptotics for Markov chains. Speeds measured for sequences $a_j$ sampled according to \eqref{e:marc} with $a=0.04$ ($\mathcal{A}$) and $a=0$ ($\mathcal{B}$). Speed plots as functions of $\mu$ (left) show square root asymptotics. Speed samples for fixed $\mu$ with varying $p$ (right) show the linear dependency on $p$ \eqref{e:markovspeed}.}
\label{fig:markov_asy}
\end{figure}

The horseshoe construction within the cat map illustrates how random dynamics occur naturally within the class of media we focus on here. Indeed, choosing characteristic functions for $A:\mathcal{M}\to\R$ on a Markov partition, we can realize quite general Markov chains. We therefore study directly media generated by Markov chains. As an example, we generated random sequences based on the Markov graph 
\begin{equation}\label{e:marc}
\mathcal{A}\stackrel{1-p}{\to} \mathcal{B} \stackrel{1}{\to} \mathcal{A}.
\end{equation} 
that is, events $\mathcal{B}$ are followed by event $\mathcal{A}$ and $\mathcal{A}$ follows $\mathcal{A}$ with probability $p$. 
% \[ 
% \begin{bmatrix}
% p & 1-p \\ 
% 1 & 0
% \end{bmatrix},\; \;
% p \in \left[ 0,1 \right] 
% \]
% we generate an ergodic Markov chain $\left \{ x_{i} \right \}$ with $x_0 = 1$, 
We define the medium $a_j = 0.05$ for event $A$ and $a_j=0$, otherwise.  We show measured speeds for $d = 0.1$
in Figure \ref{fig:markov_asy}. 
A rough heuristic would assume that the speed is determined by the medium, locally, with passage times around lattice point $T_j\sim |\mu-\mu_\mathrm{c}|^{-1/2}$. Noticing that points with almost pinning occur only in state $\mathcal{A}$, which occurs with probability $(2-p)^{-1}$ in the stationary distribution, we obtain an expected average speed 
\begin{equation}\label{e:markovspeed}
\bar{s}\sim (2-p)|\mu-\mu_\mathrm{c}|^{1/2};
\end{equation}
see Figure \ref{fig:markov_asy}.
An interesting, different class of random media is generated by a continuous distribution for the values $m_j$. We shall explore here power law distributions that emphasize different extreme value statistics. We sample $m$ from a uniform distribution and distribute the $a_j$ as $\rho\sim m^\gamma$, that is, according to a density $\rmd m=\frac{1}{\gamma}\rho^{-1+1/\gamma}\rmd \rho$. 

\begin{figure}[h!]
\includegraphics[width=\columnwidth]{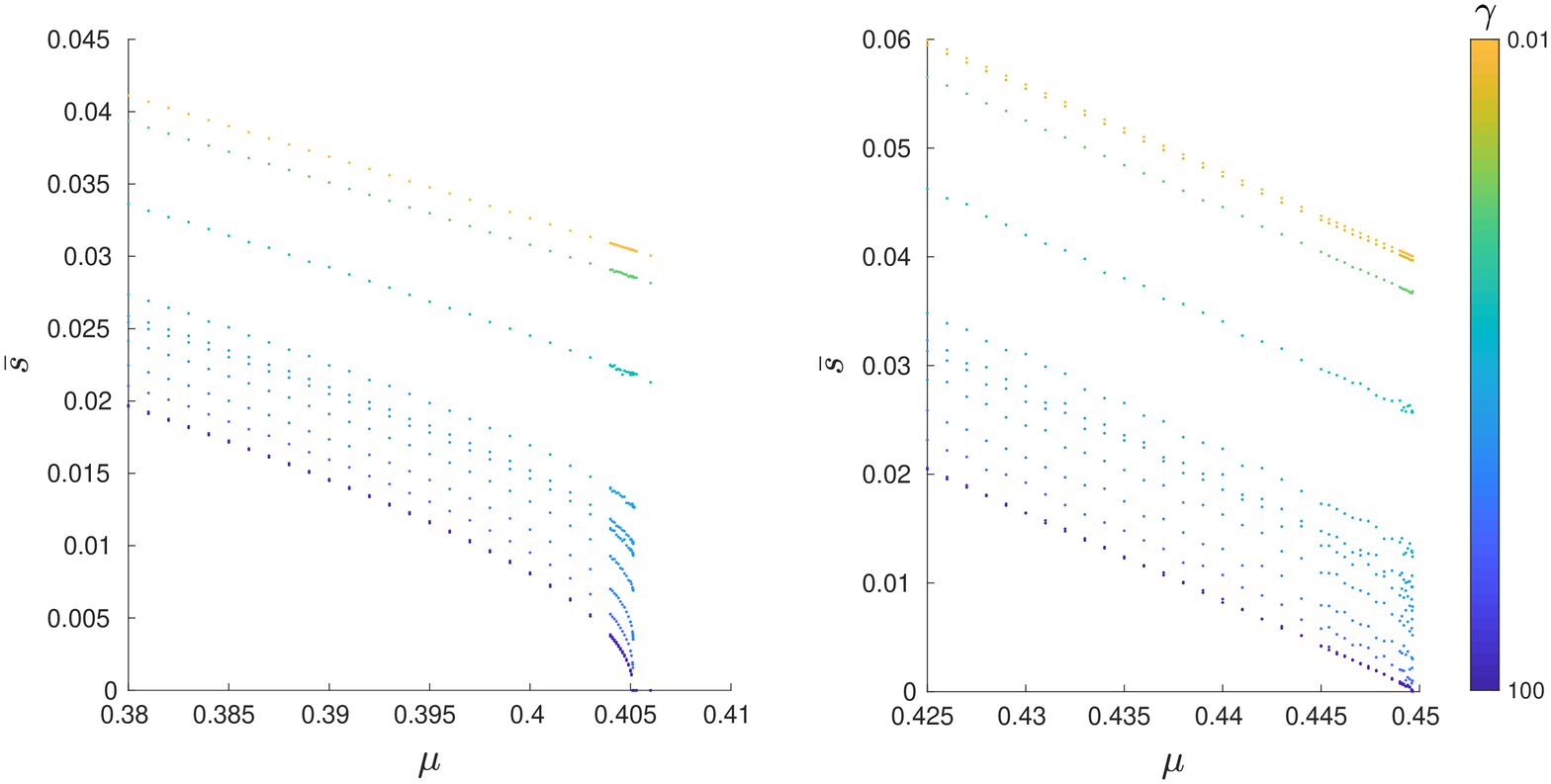}
\caption{Speeds measured for sequences $a_j$ sampled according to power-law distribution with exponent $\gamma$; smooth depinning for $d=0.1$ (left) and hard depinning for $d=0.3$ (right); see text for detail. 
}
\label{fig:rand_dist_raw_data}
\end{figure}

\begin{figure}[h!]
\includegraphics[width=\columnwidth]{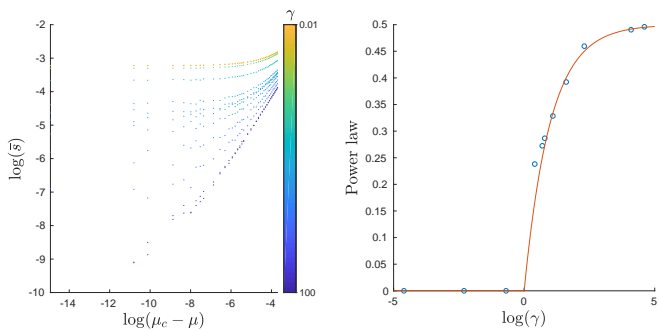}
\caption{ Power-law behavior in speeds from Figure \ref{fig:rand_dist_raw_data}, $d=0.1$ in a log-log plot near the depinning transition, for large values of $\gamma$ (right) and best-fit power laws as functions of $\gamma$.}
\label{fig:rand_dist_log_stuff}
\end{figure}

If we assume that the front speed is determined by the nearest lattice point, only, by the usual square-root depinning laws, we find passage times near a lattice point and average speeds that scale as 
\[
T_j\sim (\rho_j+|\mu-\mu_\mathrm{c}|)^{-1/2}, \qquad \bar{s}=\left(\lim_{J\to\infty} \frac{J}{\sum_{j=1}^J T_j}\right).
\]
The inverse average of the independent variables in the denominator converges to the expectation, 
\begin{align}\label{e:rand}
\bar{s}&=\left(\int  (\rho+|\mu-\mu_\mathrm{c}|)^{-1/2}\frac{1}{\gamma}\rho^{-1+1/\gamma}\rmd \rho \right)^{-1}\nonumber\\
&\sim |\mu-\mu_\mathrm{c}|^{\frac{1}{2}(1-\frac{1}{\gamma})},
\end{align}
for $\gamma>1$. The calculation is analogous to the calculation of average speeds in ergodic media and we find the effective corresponding dimension,
\[
\kappa=1/\gamma,
\]
that is, the inverse sampling exponent gives the ergodic dimension of the associated measure. Figure \ref{fig:rand_dist_raw_data} illustrates the depinning asymptotics, showing a general decrease in the speed with $\gamma$, consistent with the fact that large values of $\gamma$ increase the likelihood of almost-pinning events, and also an increase in the exponent of the power law. Figure \ref{fig:rand_dist_log_stuff} shows a comparison of the power laws with our prediction \eqref{e:rand}. 

The same calculation shows that sampling $a_j$ from a Bernoulli distribution, say $p(0)=q,\ p(0.1)=1-q$, gives depinning asymptotics,
\[
\bar{s}\sim \frac{1}{q} (|\mu-\mu_\mathrm{c}|)^{1/2}
\]
replicating \eqref{e:markovspeed}. Note that these asymptotics are steeper for rare almost-pinning events, $q\ll 1$; see Figure \ref{fig:markov_asy}.

 %%%%%%%%%%%%%%%%%%%%%
\section{\label{sec:limits}Limits of Smoothness}
We have shown thus far ample evidence that supports the assumption of a reduction to scalar positional dynamics, depending smoothly on the medium as in \eqref{e:pm}. In translation-invariant media with a weak inhomogeneity, 
\begin{equation}\label{e:mc}
u_t=u_{xx}+u(1-u)(u-\mu-\epsilon a(x;m)),
\end{equation}
where $a(x;m)$ is generated by a continuous flow $\psi^x(m)$ on a smooth manifold $\mathcal{M}$,  $a(x;m)=A\circ \psi^x(m)$, one can establish such a smooth description using perturbative methods.\cite{AScheel1} The analysis there relies, however, on an assumption of ``Lyapunov dominance'', roughly requiring that the Lyapunov exponents of the flow $\psi$ generating the medium are bounded relative to the exponential decay rate of the front solution. This suggests qualitatively different behavior for lattice-differential equations on a medium with positive Lyapunov exponents (relative to the ergodic measure relevant for the statistics of the particular realization of the medium).

Indeed, exponential convergence rates of the front solution are 
\[
u_j\sim \nu^{-|j|}, \ \nu\sim (1 + 4 d + \sqrt{1 + 8 d})/(4 d) \mbox{ for } \mu+a_j\sim 1/2,
\] 
with large-$d$ asymptotics $\nu\sim 1+1/\sqrt{2d}$. For the cat map, the unstable Lyapunov exponent is $1+\gr$ and equals the unstable exponent of the medium for $d=0.5$; the $2$:$1$-resonance occurs at $d\sim 0.15$. 

\begin{figure}[t!]
\includegraphics[width=90mm]{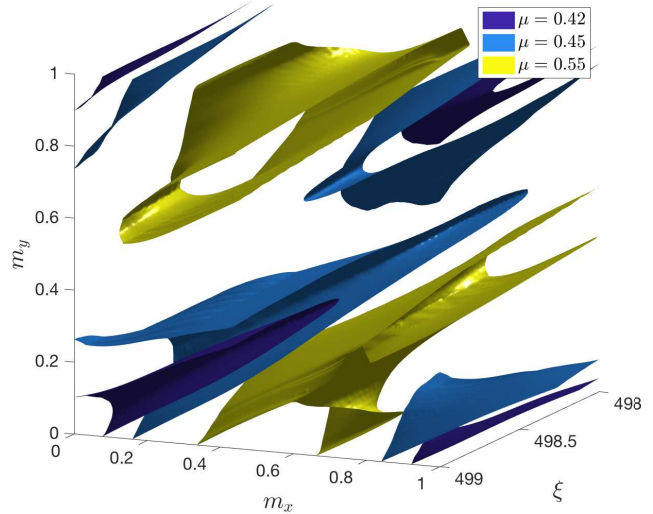}
\caption{The isosurfaces of the pinned states in the cat map medium for three values of $\mu$ with $d=0.25$; compare the strong folding of level surfaces in comparison with Figure \ref{fig:cat_iso}, which ultimately leads to irregular, difficult to resolve level surfaces when $d$ is increased further.}
\label{fig:cat_discont}
\end{figure}

\begin{figure}[t!]
\centering
\includegraphics[width=80mm,trim={2cm 0 2.5cm 2cm},clip]{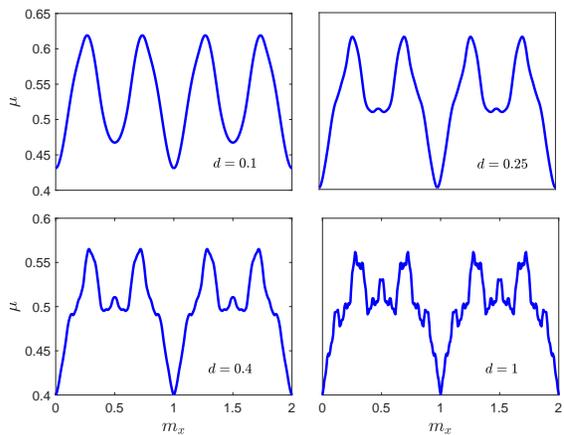}
\caption{Pinning relations in the cat map at $\xi=m_y=0$ showing $\mu_\mathrm{p}(m_x)$, periodically extended. As described in the text, pinning surfaces loose regularity as $d$ is increased.\cite{dead}}
\label{fig:cat_mu_mx}
\end{figure}

Practically, we observe this difficulty when computing pinning relations in \eqref{e:nqp} and continuing in the medium. Evaluating derivatives with respect to the medium in a Newton continuation algorithm becomes unfeasible when the derivative with respect to the medium at the front 
\begin{equation*}
\begin{split}
\partial_m& F(u_j,\xi,m,\mu)=\\
&-u_j(1-u_j)\epsilon\sin(u_j-\mu-(\psi^j(m)_x))\partial_{m_x}\psi^j(m).
\end{split}
\end{equation*}
is not spatially localized. We notice the competing effect of $u_j(1-u_j)\to 0$ with the convergence rate of the front, and of $\partial_{m_x}\psi^j(m)$ which grows with rate $1+\gr$.\footnote{One readily finds that pinning relations are H{\"o}lder continuous in $m$ for all values of $d$, for small enough H{\"o}lder exponents $\iota$, by balancing the divergence of $|\psi^j(m_1)-\psi^j(m_2)|^\iota\sim \nu^{-\iota j}$ with the convergence of the front.} Computations of the pinning relations reveal a steepening of isosurfaces as $d$ is increased, leading to irregular surfaces for larger $d$ that are difficult to resolve computationally; see Figure \ref{fig:cat_discont} for the isosurfaces and Figure \ref{fig:cat_mu_mx} for a plot of $\mu$ along a section through the medium. 

Beyond the structure of pinning surfaces, we tested the impact of this loss of smoothness on depinning asymptotics, in the example of random media. Figure \ref{fig:rand_dist_raw_data} shows the change in depinning asymptotics  when $d$ is increased from $d=0.1$ to $d=0.3$. Clearly, speeds are increased near the depinning transition and depinning asymptotics favor hard, discontinuous depinning transitions. Heuristically, this is caused by the fact that extreme events are less frequent when the front is wider and therefore interacts with multiple lattice sites at a time. Within the numerical resolution, here, depinning was discontinuous for all moderate values of $\gamma$. 

More quantitatively, assuming non-smooth behavior near the extrema of the local speed $s\sim s_\mathrm{max}+c_2(|m|+|\xi|)^q$ in \eqref{e:sasy} leads to power laws $\bar{s}\sim \tilde{\mu}^{1-\frac{\kappa+1}{q}}$ with transition to hard depinning for the smaller dimension $\kappa=q-1$. For $q<1$, depinning is always hard. 

In order to analytically explore this phenomenon, we studied what we believe to be the simplest toy example of a medium with nonzero Lyapunov exponents, a heteroclinic medium in continuous space $x\in\R$, \eqref{e:mc} with $\epsilon=0.2$ and $a(x;m_0):=\psi^x(m_0)=m(x;m_0)$, the solution to 
\begin{equation}\label{e:hetm} 
m'(x)=\tau\sin(\pi m(x)),\qquad m(0)=m_0
.\end{equation}
For $m\in(0,1)$, $a(x;m)\to 0$, $x\to -\infty$, and $a(x;m)\to 1$, $x\to +\infty$, with exponential rate $\tau\pi$. Thinking of $\epsilon=0.2$ as a small perturbation, one can compute an effective speed as a function of $x$ perturbatively, in particular near $|x|=\infty$. Projecting the effect of the medium $-\epsilon u(1-u)a(x;m)$ on the Goldstone mode given by the derivative of the front $u_\mathrm{f}'(x)\sim \rme^{-\sqrt{2}|x|}$, we find, 
\[
s(x)\sim s_\pm(\mu)+c_1\rme^{-\sqrt{2}|x|}+c_2 \rme^{-\tau\pi|x|},\qquad c_1,c_2\neq 0.
\]
Since $m(x)\sim \rme^{-\tau\pi|x|}$, we can express $s$ in terms of the medium as 
\begin{equation}\label{e:sm}
s(m)\sim s_\pm(\mu) +\left\{\begin{array}{ll}
m+\mathcal{O}(m^{\sqrt{2}/(\tau\pi)}),& \sqrt{2}>\tau\pi,\\
m^{\sqrt{2}/(\tau\pi)}+\mathcal{O}(m),& \sqrt{2}<\tau\pi.
\end{array}
\right.
\end{equation}
\begin{figure}[h!]
\includegraphics[width=.48\columnwidth]{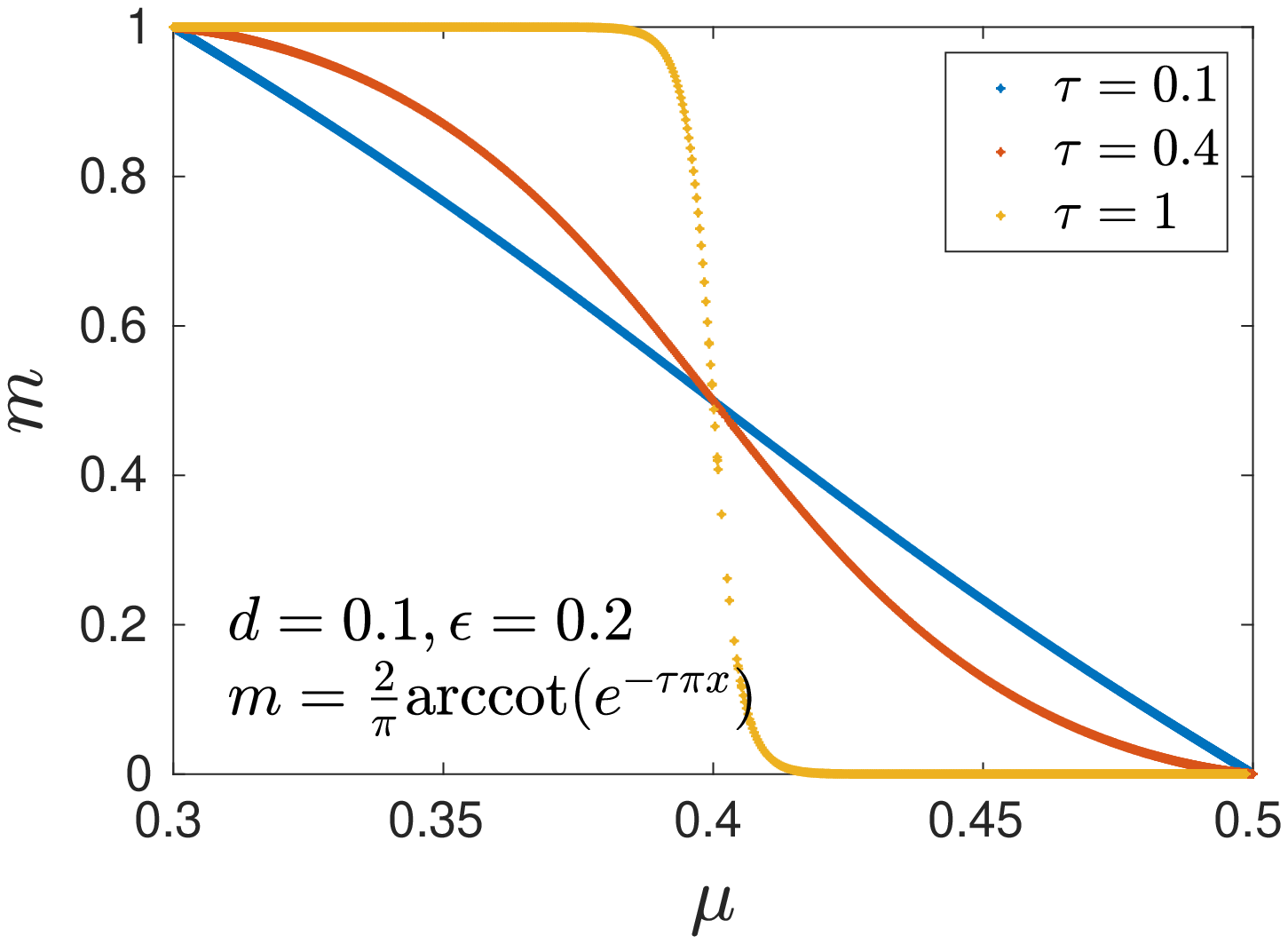}\hfill
\includegraphics[width=.48\columnwidth]{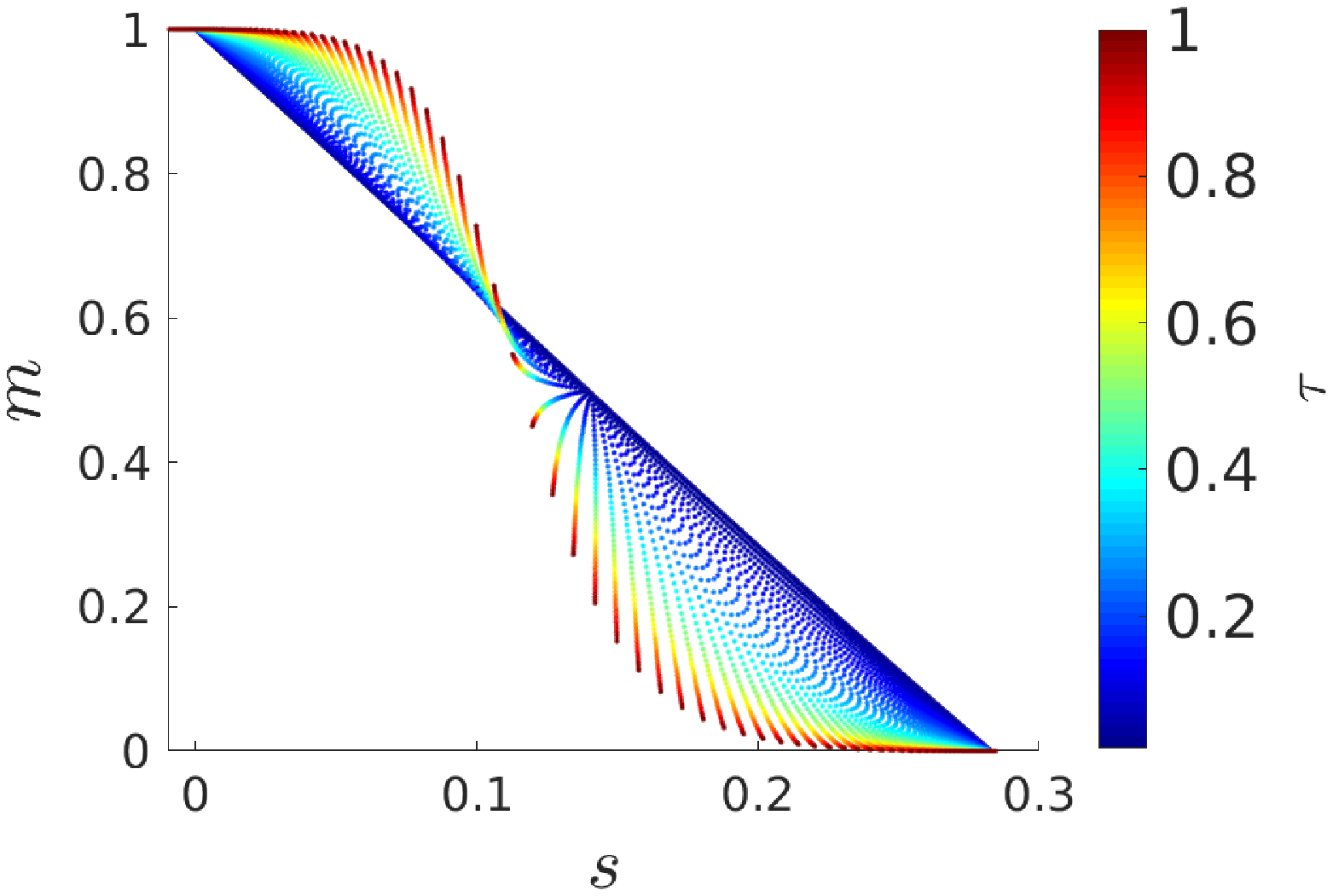}
\caption{Left: Pinning relations in the heteroclinic medium generated by \eqref{e:hetm} for convergence rates of the medium $\tau=0.1,0.4,1$.  Note that the pinning relation $\mu=\mu_\mathrm{p}(m)$ (inverse of the graph shown here) looses smoothness for larger $\tau$ near $m=0$ and $m=1$. Right: Instantaneous speeds $s(m)$ as in \eqref{e:sm}, for various values of $\tau$ (color coding); values of  $\mu$ are $\left\{ 0.1,0.3,0.5,0.7 \right\}$. Note the lack of smoothness of speed as a function of the medium for larger values of $\tau$, reflecting the lack of smoothness of pinning relations in Figure \ref{fig:tau_dep}.}
\label{fig:tau_dep}
\end{figure}
In the latter case of large Lyapunov exponents in the medium, compared to decay rates of the front, the leading order term is clearly not smooth. Figure \ref{fig:tau_dep} shows the associated dependence of pinning relations with a change to non-smooth behavior as $\tau$ increases, with a cross-over to non-smoothness predicted at the $1$:$1$-resonance $\tau\sim 0.45$. 
We also measured instantaneous speeds in the medium, which show a similar crossover; see Figure \ref{fig:tau_dep}.
% \begin{figure}[h!]
% \caption{Instantaneous speeds $s(m)$ as in \eqref{e:sm}, for various values of $\tau$ (color coding); values of  $\mu$ are $\left\{ 0.1,0.3,0.5,0.7 \right\}$. Note the lack of smoothness of speed as a function of the medium for larger values of $\tau$, reflecting the lack of smoothness of pinning relations in Figure \ref{fig:tau_dep}.}
% \label{fig:tau_speed}
% \end{figure}

\section{\label{sec:dis}Discussion} 
We presented a study of pinning relations and depinning asymptotics in media parameterized over smooth manifolds. Depinning asymptotics depend on extreme-value statistics in a universal fashion, with power law exponent $1-\kappa/2$, $\kappa$ the dimension of the ergodic measure. This complements findings in continuous media \cite{AScheel1}, where analogous power laws were found, with dimension reduced by 1 due to the lack of underlying lattice periodicity. Quite different power laws have been observed in media with nonlocal coupling \cite{anderson}, with power law exponents $3/2$ or higher, which corresponds to an ergodic dimension $\kappa\leq -2$ in our setting. In many cases, the medium is not independent of the front and one refers to ``self-pinning'' when front speeds vanish for open parameter values. An extreme case of such an interaction are fronts between periodic states and trivial states observed for instance in the Swift-Hohenberg equation \cite{pomeau,knobloch}. We also presented limits to such a description, which turns out to be effective only when Lyapunov exponents related to the modeling of the medium are bounded in terms of localization of the front. We demonstrate how this leads to both non-smooth pinning relations and changed depinning asymptotics. 
% We hope that the point of view of generating media through smooth dynamical processes will help to put transitions in more complex, possibly time-dependent media in a new light and allow for more systematic descriptions.

% -- singularities and bifurcation --- cite yip?


\providecommand{\noopsort}[1]{}\providecommand{\singleletter}[1]{#1}%
%

% \nocite{*}
% \bibliography{pinning}% Produces the bibliography via BibTeX.

\begin{thebibliography}{15}
\bibitem [{dea()}]{dead}%
  \BibitemOpen
  \href@noop {} {\bibinfo  {journal} {Any resemblance in the figure to actual
  cats, living or dead, is purely coincidental}\ }\BibitemShut {NoStop}%
\end{thebibliography}

\end{document}